 \documentclass[final,5p,times,twocolumn]{elsarticle}

\usepackage{amssymb}

\usepackage{lipsum}

\journal{International Journal of Mass Spectrometry}

\begin{document}

\begin{frontmatter}

\title{Collision-Induced Dissociation at TRIUMF's Ion Trap for Atomic and Nuclear Science}

\author[Triumf,UBC]{A. Jacobs\corref{cor}}
\ead{ajacobs@triumf.ca}

\author[SFU]{C. Andreoiu}

\author[JLU,GSI]{J. Bergmann}

\author[McGill]{T. Brunner}

\author[JLU,GSI]{T. Dickel}

\author[Triumf,UVic]{I. Dillmann}

\author[Triumf,York]{E. Dunling}

\author[Calgary]{J. Flowerdew}

\author[Triumf]{L. Graham}

\author[Manitoba]{G. Gwinner}

\author[Triumf,McGill]{Z. Hockenbery}

\author[Triumf,Manitoba]{B. Kootte}

\author[Triumf,UBC]{Y. Lan}

\author[Mines]{K. G. Leach}

\author[Triumf,UBC]{E. Leistenschneider}

\author[Triumf,UBC]{E.M. Lykiardopoulou}

\author[York]{V. Monier}

\author[Triumf]{I. Mukul}

\author[Triumf,Heidelberg]{S.F. Paul}

\author[JLU,GSI]{W.R. Pla{\ss}}

\author[Triumf,JLU,Edin]{M.P. Reiter}

\author[JLU,GSI,FAIR]{C.Scheidenberger}

\author[Calgary]{R. Thompson}

\author[Triumf]{J.L Tracy}

\author[JLU]{C. Will}

\author[Calgary]{M.E. Wieser}

\author[IAI]{M. Yavor}

\author[Triumf,UBC]{J. Dilling}

\author[Triumf,UVic]{A.A. Kwiatkowski}

\address[Triumf]{TRIUMF, 4004 Wesbrook Mall, Vancouver, British Columbia V6T 2T3, Canada}

\address[UBC]{Department of Physics and Astronomy, University of British Columbia, Vancouver, British Columbia V6T 1Z1, Canada}

\address[SFU]{Department of Chemistry, Simon Fraser University, Burnaby, British Columbia V5A 1S6, Canada}

\address[JLU]{II. Physikalisches Institut, Justus-Liebig-Universit\"{a}t, 35392 Gie{\ss}en, Germany}

\address[GSI]{GSI Helmholtzzentrum f\"{u}r Schwerionenforschung GmbH, Plankstra{\ss}e 1, 64291 Darmstadt, Germany}

\address[McGill]{Physics Department, McGill University, H3A 2T8 Montr\'{e}al, Qu\'{e}bec, Canada}

\address[UVic]{Department of Physics and Astronomy, University of Victoria, Victoria, British Columbia V8P 5C2, Canada}

\address[York]{Department of Physics, University of York, York YO10 5DD, United Kingdom}

\address[Calgary]{Department of Physics and Astronomy, University of Calgary, Calgary, Alberta T2N 1N4}

\address[Manitoba]{Department of Physics and Astronomy, University of Manitoba, Winnipeg, Manitoba R3T 2N2, Canada}

\address[Mines]{Department of Physics, Colorado School of Mines, Golden, Colorado 80401, USA}

\address[Heidelberg]{Ruprecht-Karls-Universit\"{a}t Heidelberg, D-69117 Heidelberg, Germany}

\address[Edin]{Institute for Particle and Nuclear Physics, University of Edinburgh, Edinburgh EH9 3FD, United Kingdom}

\address[FAIR]{Helmholtz Forschungsakademie Hessen f{\"u}r FAIR (HFHF), 35392 Gie{\ss}en, Germany}

\address[IAI]{Institute for Analytical Instrumentation, Russian Academy of Sciences, 190103 St. Petersburg, Russia}

\cortext[cor]{Corresponding author: 4004 Wesbrook Mall, Vancouver, BC V6T 2T3, Canada}

\begin{abstract}

The performance of high-precision mass spectrometry of radioactive isotopes can often be hindered by large amounts of contamination, including molecular species, stemming from the production of the radioactive beam. In this paper, we report on the development of Collision-Induced Dissociation (CID) as a means of background reduction for experiments at TRIUMF's Ion Trap for Atomic and Nuclear science (TITAN). This study was conducted to characterize the quality and purity of radioactive ion beams and the reduction of molecular contaminants to allow for mass measurements of radioactive isotopes to be done further from nuclear stability. This is the first demonstration of CID at an ISOL-type radioactive ion beam facility, and it is shown that molecular contamination can be reduced up to an order of magnitude.

\end{abstract}

\begin{keyword}

Mass Spectrometry
\sep MR-TOF-MS
\sep TITAN
\sep Radioactive Ion Beams
\sep Collision Induced Dissociation

\end{keyword}

\end{frontmatter}

\section{Introduction}

In all fields of experimental physics, one of the principal concerns is the reduction of background and the optimization of signal-to-noise. In particular, high-precision experiments, which utilize radioactive nuclei, normally require pure samples of the isotope of interest. Furthermore, if the experiment involves a Radioactive Ion Beam (RIB), the preferred beam composition is predominantly of the isotope of interest.

For such experiments, the RIB is typically produced using accelerator facilities. The method of Isotope Separation On-Line (ISOL) \cite{Blumenfeld2013FacilitiesProduction} produces RIB by impinging light projectiles (e.g. protons) on a thick target. Through the ensuing spallation, fragmentation, and fission reactions in the target material, a wide variety of isotopes are produced. The ISOL method delivers RIB to subsequent experiments with both excellent beam properties (i.e. small relative longitudinal and transverse energy spread) and at high intensities. However, one key drawback to this method is the presence of unwanted stable and radioactive contamination, including molecules. 

The alternative method for RIB production is known as the in-flight technique \cite{Blumenfeld2013FacilitiesProduction}. With this method, a high-energy beam is produced from a thin target which avoids the chemical effects of the ISOL technique. For low-energy experiments, these beams must be decelerated in gas stopping cells, where molecular contamination can be introduced. In some instances, the isotope of interest will even be measured as a molecular ion \cite{Hamaker2021}. Thus, there is a need for the reduction of molecular contamination regardless of the RIB production method.

Due to contamination always being an issue posed to ion trapping, several techniques for RIB purification already exist. For a Penning trap, individual species can be removed through the process of dipole cleaning \cite{Blaum2004}. Additionally, the method of Stored Waveform Inverse Fourier Transform (SWIFT) can provide broadband cleaning in a Penning trap \cite{KWIATKOWSKI20159}. Lastly, the method of mass-selective re-trapping using a Multiple-Reflection Time-of-Flight Mass-Spectrometer allows for fast, high-resolution, and high efficiency mass separation \cite{Dickel2017IsobarRe-Trapping}. 

With molecular contamination presenting an issue regardless of RIB production method and over a wide range in mass, attempts have been made to specifically reduce the amount of molecular contamination present in experimental setups. For example, the use of an RF carpet to transport atomic and molecular ions allows for their separation based on ion mobility \cite{MISKUN2021116450}. Additionally, breaking the bonds of molecular ions such that the ion is no longer isobaric has been shown to be a promising technique. This process, called Collision-Induced Dissociation, has previously been employed at two different gas stopping cell facilities \cite{Schury2006BeamCell, Greiner2020RemovalMethod}. However, this technique has not yet been demonstrated at an ISOL-type RIB facility.

At TRIUMF's Ion Trap for Atomic and Nuclear science (TITAN) \cite{Dilling2003TheIsotopes}, high-precision experiments are carried out for studies of nuclear structure, nuclear astrophysics, and fundamental symmetries. For these experiments, pure samples are needed to reduce the effect of ion-ion interaction as well as to avoid the limiting dynamic range (i.e. the ratio of the species of interest to contamination). At TITAN, there are two main techniques for beam purification. The first technique is isobaric separation using the recently established method of mass-selective re-trapping in a Multiple-Reflection Time-of-Flight Mass-Spectrometer  \cite{Dickel2017IsobarRe-Trapping}, which will be briefly discussed in section 2. The second technique, Collision-Induced Dissociation (CID), has been implemented to reduce loosely bound molecular contamination, the results of which are reported in this paper. This technique will be discussed in section 3 and an experimental characterization in section 4.

\section{The TITAN Experiment}

TITAN is located at TRIUMF's Isotope Separator and ACcelerator (ISAC) facility \cite{Ball2016TheBeams} in Vancouver, Canada. Radioactive Ion Beam (RIB) is produced by a primary $\approx$ 500 MeV proton beam impinging on a variety of thick targets. The nuclear reaction products diffuse out of the target and are ionized. The RIB is then transferred through a dipole magnet mass separator and delivered to the TITAN experiment.

\begin{figure}[h]
    \centering
    \includegraphics[width= 0.45\textwidth]{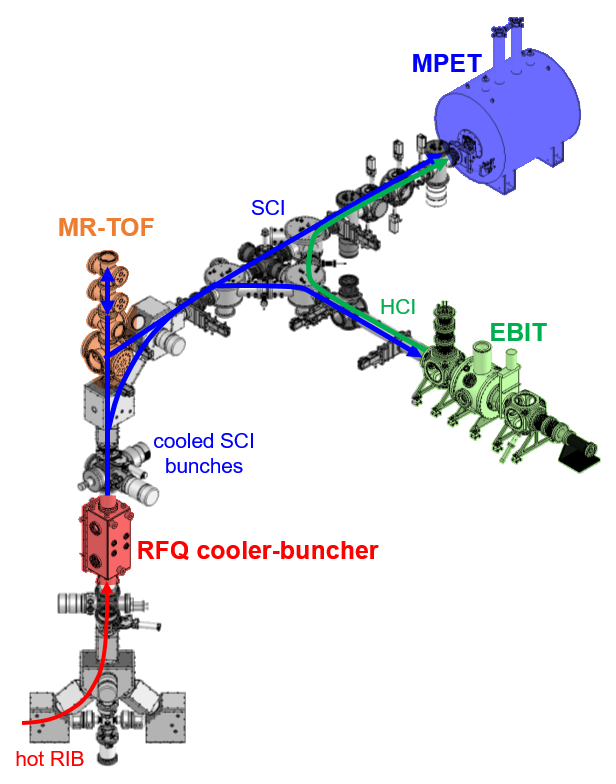}
    \caption{The current TITAN experiment \cite{Dilling2003TheIsotopes} with all traps labelled . Continuous beam is delivered from TRIUMF's Isotope Separator and ACcelerator (ISAC) to the RFQ Cooler-Buncher. The beam is then sent to the various other traps for preparation and/or measurement. CID and is performed using the MR-TOF-MS.}
    \label{fig:TITAN}
\end{figure}

The TITAN experiment \cite{Dilling2003TheIsotopes} is a series of ion traps that can be combined in different configurations to conduct high-precision mass spectrometry. An overview of the experimental set-up can be seen in Figure \ref{fig:TITAN}. The two relevant traps used during this experiment are the Radio-Frequency Quadrupole (RFQ) Cooler-Buncher and the Multiple-Reflection Time-of-Flight Mass Spectrometer.

The RFQ Cooler-Buncher is a He-gas-filled trap that receives continuous RIB from ISAC \cite{Brunner2012TITANsPerformance}. The trap then cools and bunches the beam before sending the cold ion bunches to the subsequent traps. The most recent addition to the TITAN experiment is the Multiple-Reflection Time-of-Flight Mass Spectrometer (MR-TOF-MS) \cite{Jesch2015TheTRIUMF}. This device can be used for precision mass spectrometry with relative uncertainties on the order of 10$^{-7}$ \cite{Leistenschneider2018DawningIsotopes,  Reiter2018QuenchingIsotopes}. Because of its high sensitivity, the MR-TOF-MS is particularly useful for measurements of low-yield species \cite{Beck2021MassShell,Mukul2021,Izzo2021}. This allows the TITAN MR-TOF-MS to be employed as a beam diagnostic tool for identifying the composition of a RIB \cite{Reiter2020ImprovedMR-TOF-MS}. Additionally, the MR-TOF-MS can be used to isobarically clean RIB for delivery to the rest of TITAN using mass-selective re-trapping (as described in Section 2.1 as well as in \cite{Beck2021MassShell}). The MR-TOF-MS is where CID is performed and analyzed.

\subsection{The Multiple-Reflection Time-of-Flight Mass Spectrometer (MR-TOF-MS)}

The TITAN MR-TOF-MS was designed and built at JLU-Gie{\ss}en \cite{Jesch2015TheTRIUMF} and is based on the design of the MR-TOF-MS at GSI Darmstadt \cite{Pla2008IsobarFacilities,Dickel2015ANuclei}. The device consists of a series of He-gas-filled RFQs, RF switchyard \cite{Pla2015High-performanceSpectrometry}, and RF injection trap and a time-of-flight analyzer with two electrostatic ion mirrors \cite{Yavor2015Ion-opticalSeparator}. Bunched beam arrives from the RFQ Cooler-Buncher at approximately 1.4 kV, and the ions are dynamically captured in the Input RFQ of the MR-TOF-MS. The MR-TOF-MS RFQs are driven by home built resonant drivers \cite{AndresThesis} which operate between 1 and 1.6 MHz depending on the beamline section. Input voltages were selected to provide a pseudo potential between 15 and 30 eV. In this particular experiment, settings were chosen such that the low mass cutoff was below A/q = 60 so that a wide range of masses were transported efficiently. After the ion bunch is captured in the Input RFQ, it is then re-cooled for efficient transport and transported to the injection trap via a series of RFQs. A final cooling stage is employed to ensure minimal broadening of the time-of-flight peak due to thermal energy spread before injecting the ion bunch into the time-of-flight analyzer. 

After injecting ions into the time-of-flight analyzer, the MR-TOF-MS allows for two modes of operation which results in the ions exiting from either the top or the bottom of the time-of-flight analyzer. To make the two modes of operation independent of each other, the ions undergo a time focus shift (TFS) reflection in each electrostatic mirror at the beginning of each cycle \cite{Dickel2017DynamicalSpectrometers}. The TFS results in the ion bunch having correctly located time foci on both ends of the time-of-flight analyzer. The ions then undergo a series of isochronous reflections to preserve the time foci. After the ions have flown for enough time and, as a result, spread out into isobaric bunches, one of the two aforementioned options for operation is used. 

In mass spectroscopy mode, the potentials on the electrodes on the detector side of the analyzer, the second electrostatic ion mirror, are lowered. The ions are then extracted from the analyzer and detected by a fast time-of-flight (TOF) detector (ETP MagneTOF DM572). The TOF is then recorded with a time-to-digital converter (FastCom TDC MCX6A). This generates a TOF spectrum which is used to derive the masses of the species in the spectrum. 

The other mode of operation is isobaric separation using a technique called mass-selective re-trapping \cite{Dickel2017IsobarRe-Trapping, Jesch2015TheTRIUMF}. In this mode, the potentials of the electrodes on the injection trap side of the analyzer, the first electrostatic ion mirror, are lowered. Ions then reverse back into the open injection trap. The trap, when closed at the optimal time for the species of interest, adibatically captures the ions turning around in the trap. Outside of this narrow time window ($\approx$30 ns), the other species are not captured and are discarded. After re-trapping is completed, the ions are re-cooled and either re-injected into the analyzer for mass spectrometry, or they are ejected from the MR-TOF-MS to the subsequent traps in the TITAN system as an isobarically cleaned beam. As a result, the MR-TOF-MS can act as its own isobar separator before performing mass spectrometry. Mass selective re-trapping was successfully implemented during prior experiments \cite{Beck2021MassShell}, and is now considered a well established technique \cite{Mukul2021,Izzo2021,Paul2021,Porter2022}. For a comprehensive technical review of the TITAN MR-TOF-MS, see \cite{REITER2021165823}.

\section{Collision-Induced Dissociation}

Collision-Induced Dissociation (CID) is a well established experimental technique in analytical chemistry for conducting thermodynamic experiments on molecular ions \cite{McLuckey1992PrinciplesSpectrometry}. The technique involves exciting rotational-vibrational states in molecular ions through collisions with a neutral gas. There have been a variety of experiments and applications using CID such as high-energy MS-MS \cite{Beynon1973DesignSpectrometer}, low-energy Quadrupole-Quadrupole-Time-of-Flight \cite{Morris1996HighSpectrometer} and low-energy Quadrupole Ion Trap (also called resonance excitation) \cite{Louris1987InstrumentationSpectrometry}. High-energy CID employs beams with an energy of 1 to 10 keV while low-energy CID utilizes beams of 1-100 eV.  For a more comprehensive review of CID in analytical chemistry, please see \cite{Wells2005Collision-inducedProteins}.

In nuclear physics research, CID has been used for beam purification after extraction from gas-filled stopping cells at both MSU \cite{Schury2006BeamCell} and GSI-Darmstadt \cite{Greiner2020RemovalMethod}. To perform CID in these low-energy experiments, the high energy RIB is stopped in a gas stopping cell and is transported through an RFQ mass filter before being accelerated into a gas target for CID. A final isolation stage is performed via a time-of-flight mass filter. However, CID has yet to be applied at an ISOL facility. 

\begin{figure}[t]
    \centering
    \includegraphics[width = 0.48\textwidth]{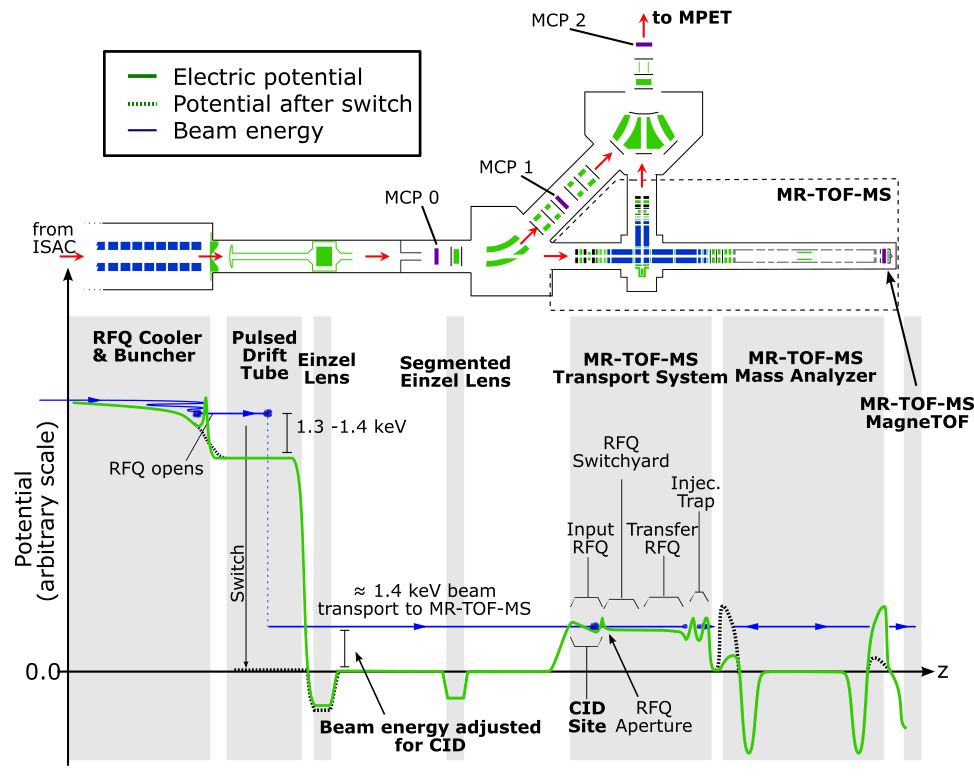}
    \caption{A schematic view of the TITAN beamline from the RFQ Cooler-Buncher to the MR-TOF-MS. The progression of the beam energy as it travels through the TITAN setup is depicted along with the electric potentials before and after switching. By varying the potential change in the Pulsed Drift Tube, the injection energy into the MR-TOF-MS can be manipulated. CID occurs in the He buffer gas filled Input RFQ of the MR-TOF-MS.}
    \label{fig:OLISTITAN}
\end{figure}

\subsection{CID with the TITAN MR-TOF-MS}

At TITAN, low-energy (10 - 110 eV) CID is implemented in a method similar to that of MSU \cite{Schury2006BeamCell} and GSI-Darmstadt \cite{Greiner2020RemovalMethod}. For systematic studies of CID, an ion cocktail of various atomic and molecular species was created in a plasma ion source at ISAC's Off-Line Ion Source (OLIS). The beam was then purified to a single mass unit in a magnetic dipole mass-separator before being sent to TITAN's RFQ Cooler-Buncher. The 20 keV bunched beam was then ejected towards the MR-TOF-MS. While traveling through a pulsed drift tube, the beam energy was reduced to approximately 1.4 keV transport energy. The beam then entered the He gas-filled Input RFQ. The potential along the RFQ segment was 1.35 kV while the input aperture was switched between 1.35 and 1.5 kV to confine the ions longitudinally. Radial confinement was provided by an $\approx$ 20 V pseudopotential from the RFQ fields. After being captured, a 6 V DC gradient transported the ions along the RFQ segment. The ions were cooled for $\approx$ 1 ms before being sent further along the beamline. By changing the potential of the pulsed drift tube, relative to the Cooler-Buncher, the energy at which the ions entered the Input RFQ of the MR-TOF-MS was precisely controlled. As a result, CID was studied by injecting molecular ions with higher energies into the Input RFQ. Figure \ref{fig:OLISTITAN} presents a schematic of the beam path along with a corresponding plot of beam energy.

\begin{figure}[t]
    \centering
    \includegraphics[width = 0.48\textwidth]{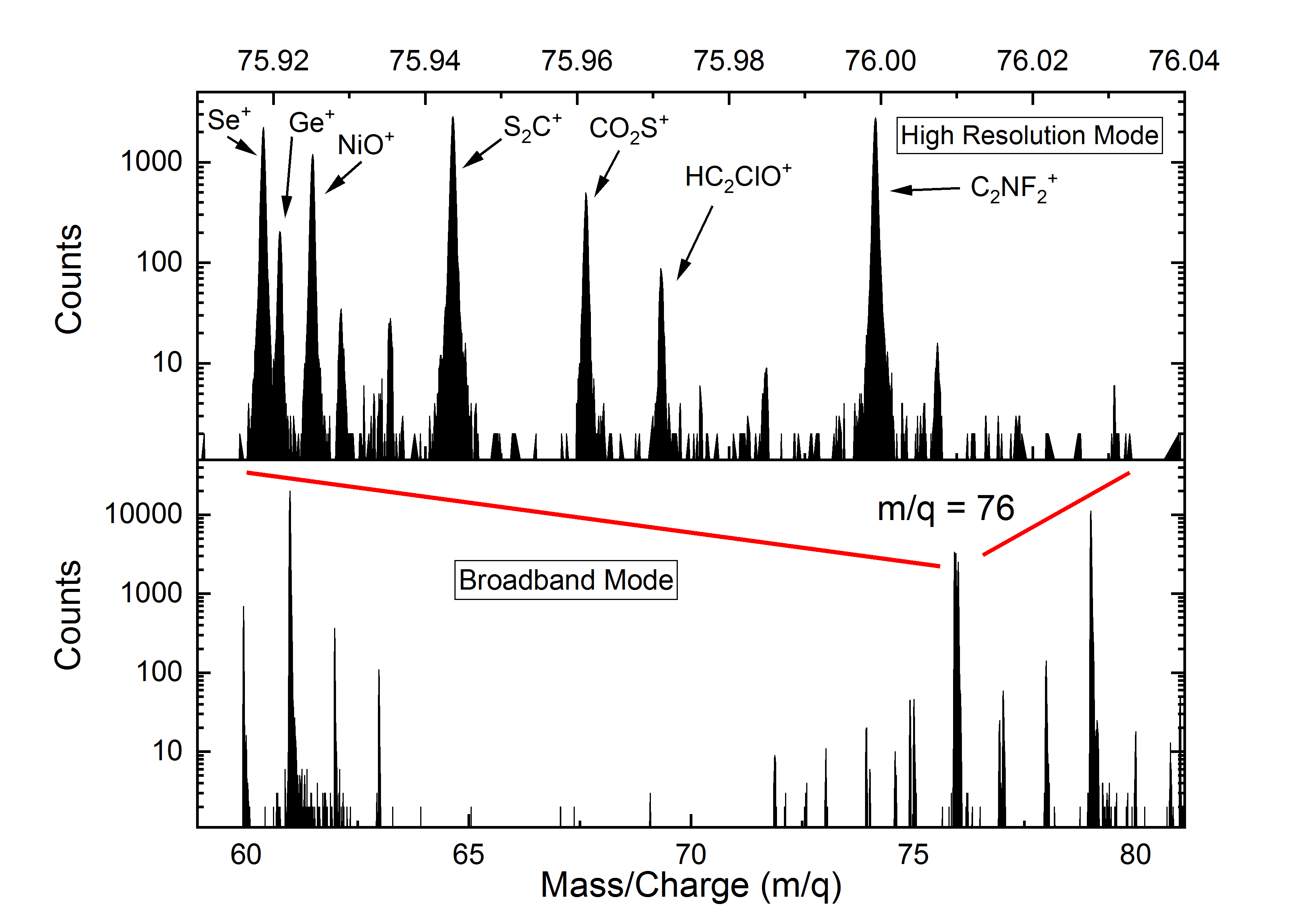}
    \caption{The bottom spectra is a broadband measurement of species seen after injecting A = 76 beam into the MR-TOF-MS. This allows for the identification of the main routes of CID and molecular addition reactions (species with masses greater than 76). The top spectrum is a high-resolution measurement focusing only on mass 76. This allows for the identification of individual atomic and molecular species found in the A = 76 beam. Data for both spectra was taken with an injection energy of 80 eV.}
    \label{fig:BroadHighRes}
\end{figure}

After the molecular ions undergo CID, they are transported to the time-of-flight analyzer section of the MR-TOF-MS. There are two different modes of MS that were utilized. 1.) Broadband mode is operated with a flight path of three isochronous turns or less. In broadband mode, a wide range of masses were measured at the same time (in Figure \ref{fig:BroadHighRes}, A/q = 60-80 u). This facilitated the measurement of the initial molecular species and the CID product species simultaneously. 2.) The high resolution mode used a flight path of at least 300 isochronous turns (total TOF $>$ 5 ms). This mode focused on a single mass unit and allowed for the identification of individual atomic and molecular species. For a comparison of the two MS modes, see Figure \ref{fig:BroadHighRes} (top and bottom for high resolution and broadband respectively).

For the characterization of CID, two different isobars were selected to be delivered by OLIS. A/q = 78 was selected as a simple spectrum with only $^{78}$Kr$^{+}$ and CH$_{2}$O$_{2}$S$^{+}$ ions, and A/q = 76 was selected as a complex spectrum with many atomic and molecular species. Initial broadband spectra were taken for each mass unit while scanning injection energy to identify the most prominent product mass units. After this, high-resolution spectra were taken at relevant mass units, established by the initial broadband spectra, while scanning injection energy to identify the breakup and production rates for individual atomic and molecular species.

\section{CID Results}

\begin{figure}[!t]
    \centering
    \includegraphics[width = 0.4\textwidth]{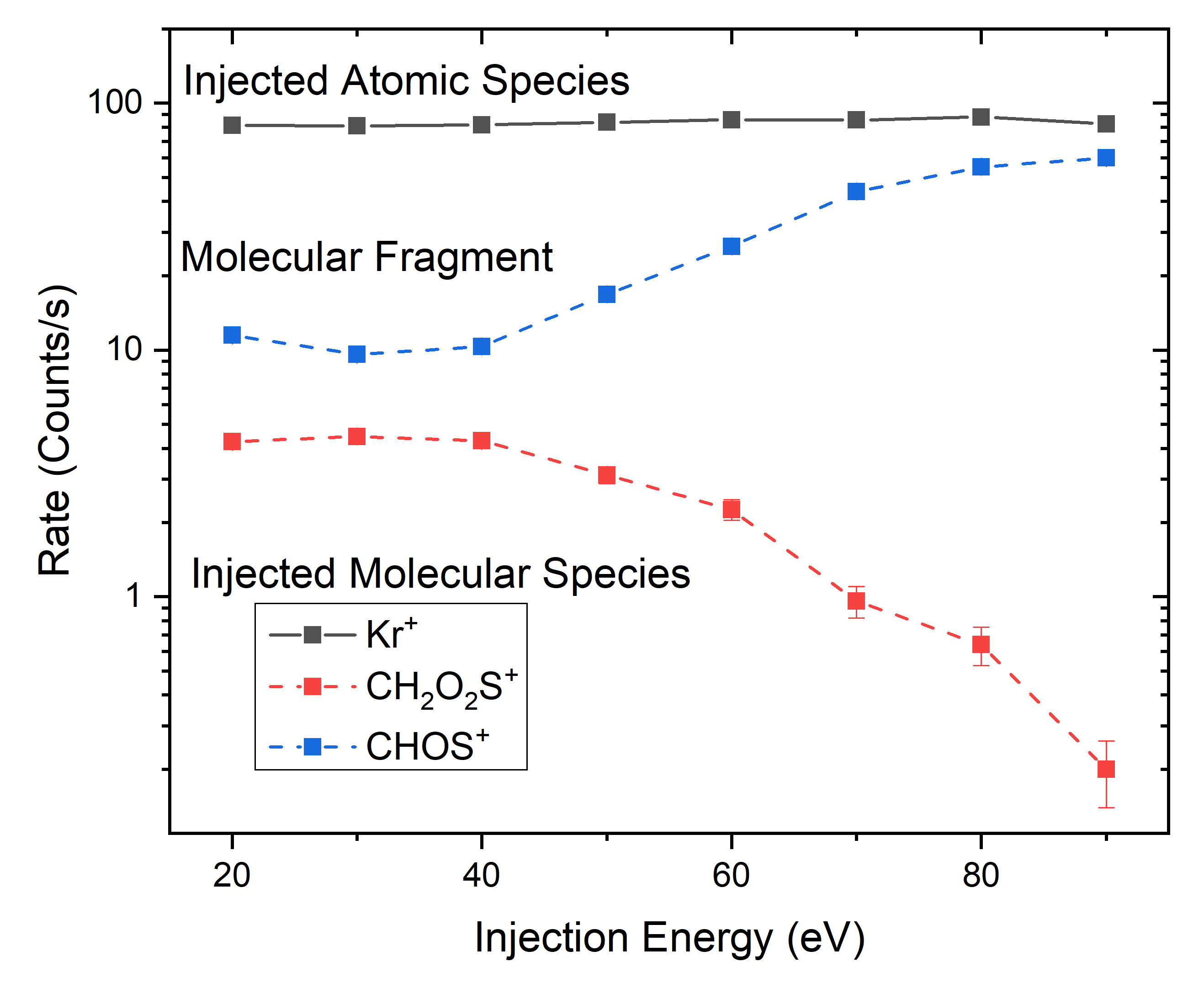}
    \caption{The rates of $^{78}$Kr$^{+}$, $\textrm{C}\textrm{H}_{2} \textrm{O}_{2} \textrm{S}^{+}$, and CHOS$^{+}$ were measured as the injection energy into the Input RFQ of the MR-TOF-MS was varied. The atomic species is shown with solid lines while molecular species are shown with dashed lines.}
    \label{fig:78A}
\end{figure}

The atomic species in the A/q = 78 beam was $^{78}$Kr$^{+}$, specifically injected into the plasma ion source, and the molecular species was identified as $\textrm{C}\textrm{H}_{2} \textrm{O}_{2} \textrm{S}^{+}$. Initially, a broadband measurement was taken to identify the strongest channels of both CID as well as molecular addition. In Figure \ref{fig:78A}, the counting rates of $^{78}$Kr$^{+}$, CH$_{2}$O$_{2}$S$^{+}$, and CHOS$^{+}$ were measured in high resolution mode (at A/q = 78 and 61 respectively) while adjusting the injection energy. As the injection energy into the Input RFQ of the MR-TOF-MS is increased, there is little to no change in the counting rate of the atomic $^{78}$Kr$^{+}$. Conversely, there is a steady decline in the counting rate of the CH$_{2}$O$_{2}$S$^{+}$ molecular ion and an increase in one of its possible breakup products CHOS$^{+}$. Additionally, for injection energies exceeding $\approx$ 100 eV, it was found that losses in the $^{78}$Kr$^{+}$ occurred due to ions' energy exceeding the potential provided by the confining aperture in the Input RFQ.

In regards to possible discrepancies seen in Figure \ref{fig:78A}, it is worth noting that, even at low injection energies, the molecular fragment's counting rate exceeds that of the injected A = 78 molecule. This is due to chemical effects (i.e. molecular addition/breakup reactions) occurring in the TITAN RFQ Cooler-Buncher before the beam is injected into the MR-TOF-MS. Additionally, it can be seen that the sum of injected CH$_{2}$O$_{2}$S$^{+}$ and breakup product CHOS$^{+}$ does not remain constant as injection energy increases. Moreover, the increase in the rate of CHOS$^{+}$ increased beyond the initial rate of CH$_{2}$O$_{2}$S$^{+}$. The reason for these apparent discrepancies is due to the breakup of CH$_{3}$O$_{2}$S$^{+}$ at A/q = 79 which had formed in the TITAN RFQ Cooler-Buncher before injection into the MR-TOF-MS. While the presence of these molecular sidebands could be a cause for concern regarding the purity of the He buffer gas, other tests with atomic ions as discussed in \cite{REITER2021165823} show a minimal effect on overall efficiency of the system. As such, the key takeaway remains that molecular breakup can be increased by a moderate increase in injection energy while atomic ions are not lost due to scattering effects.

\begin{figure}
    \centering
    \includegraphics[width = 0.42\textwidth]{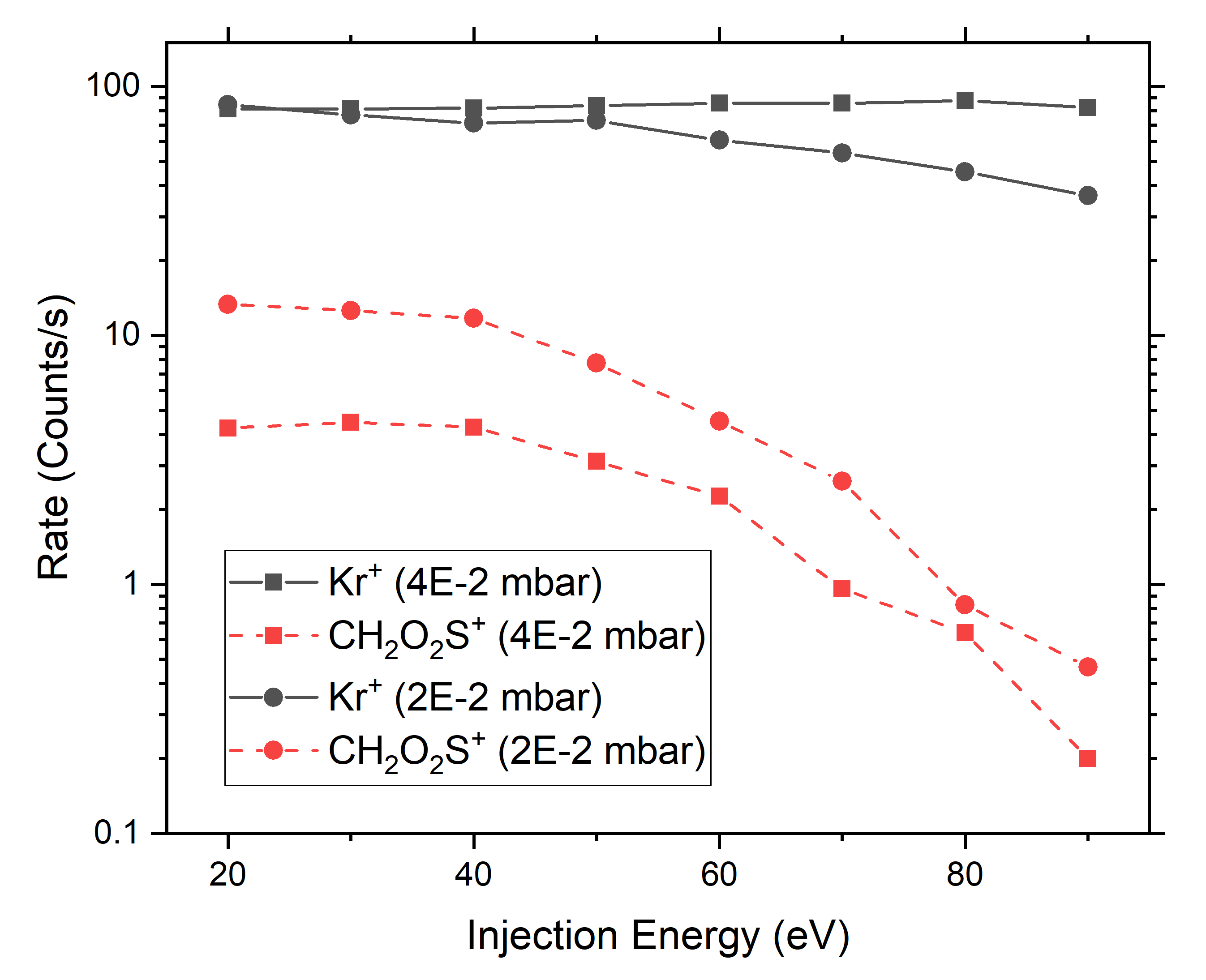}
    \caption{The comparison of atomic (black and solid lines) and molecular (red and dashed lines) transmission rates for He gas pressures of $4 \times 10^{-2}$ mbar (squares) $2 \times 10^{-2}$ mbar (circles) in the MR-TOF-MS Input RFQ.}
    \label{fig:pressures}
\end{figure}

Due to some He buffer gas travelling through the differential pumping apertures from the RFQs to the analyzer section of the MR-TOF-MS, it is important to pick a pressure that allows for both efficient capture of ions in the Input RFQ and long flight paths in the analyzer. In light of this, CID was tested using two He gas pressures in Figure \ref{fig:pressures}. The first observation is the drop in the rate of $^{78}$Kr$^{+}$ at low injection energies for a lower He pressure. The other key observation is the higher rate of CH$_{2}$O$_{2}$S$^{+}$ for the lower He pressure. Thus, it is clear that the higher He pressure is better for both the efficient capture of atomic ions as well as the breakup of molecular ions.

\begin{figure}
    \includegraphics[width = 0.4\textwidth]{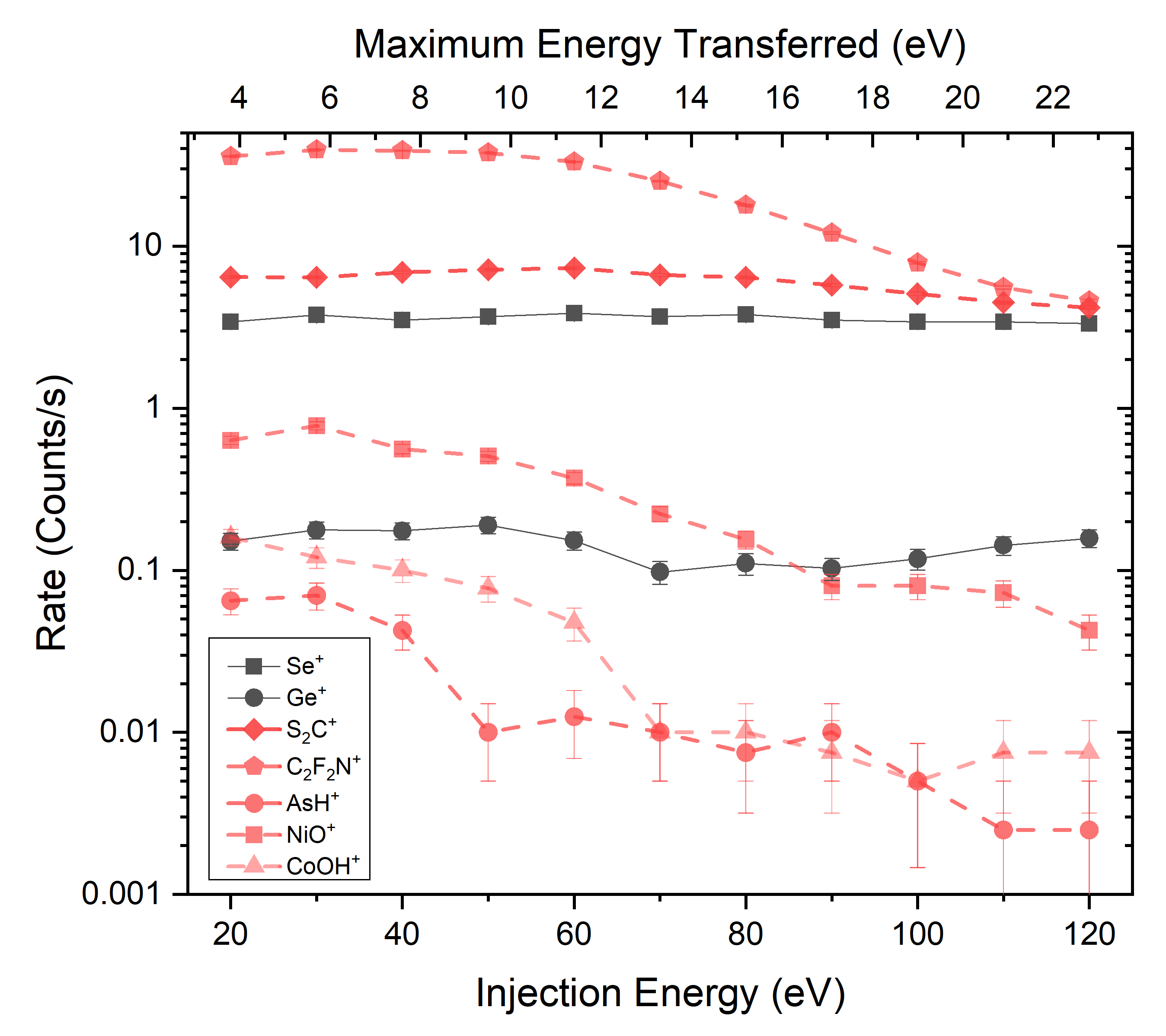}
    \caption{The rates of both atomic (black and solid lines) and molecular (red and dashed lines) ions with injection varied. The approximate maximum energy transferred is shown on the upper x-axis as calculated using equation \ref{energy_transferred}. Additionally, the bond energy of the molecular species is scaled approximately by shade.}
    \label{fig:76A}
\end{figure}

For the A/q = 76 beam, many different molecular and atomic species were identified as can be seen in the upper part of Figure \ref{fig:BroadHighRes}. In Figure \ref{fig:76A}, injection energy was scanned and counting rate was recorded for each species. In addition to injection energy, the maximum energy transferred can be  calculated using the assumption of a purely inelastic collisions and assuming hard-core point-like molecular ions:

\begin{equation}
\label{energy_transferred}
    Q_{max} = E_{rel} = \frac{m_{He}}{m_{ion} + m_{He}} E_{inj} \label{eq:etran}
\end{equation}

\noindent
where $Q_{max}$ is the maximum transferred energy, $E_{rel}$ is the relative energy between the collisional partners, $m_{He}$ is the mass of He, $m_{ion}$ is the mass of the injection ion, and $E_{inj}$ is the injection energy 
\cite{McLuckey1992PrinciplesSpectrometry}. However, this is a highly idealized treatment of the physical process of CID. A more accurate calculation uses the binary limit \cite{Boyd1984Angle-DependenceCollisions,Douglas1982MechanismSpectrometry} such that:

\begin{equation}
    Q_{max} = \frac{m_{He}}{m_{ion} + m_{He}(m_{col}/m_{ion})} E_{inj},
\end{equation}

\noindent
where $m_{col}$ is the mass of the portion of the molecule which undergoes the collision assuming an asymmetric structure. Because the TITAN MR-TOF-MS is not equipped to measure necessary parameters for collisional dynamics (e.g. angular dependencies), Equation \ref{eq:etran} is used with Figure \ref{fig:76A}. As with the $^{78}$Kr$^{+}$, it can be seen that the atomic ions ($^{76}$Se$^{+}$ and $^{76}$Ge$^{+}$) are unaffected by the increase in injection energy. Additionally, it can be seen that not all molecular species declined the same amount or with the same trend. This indicates that the process of CID is species dependent, as expected. Furthermore, as can be seen in both Figures \ref{fig:78A} and \ref{fig:76A}, the transmission of molecular ions can be reduced by up to an order of magnitude. This amount of reduction can prove crucial for the measurement of low rate species found in RIB that would be otherwise obscured by molecular contamination.

\begin{figure}[!b]
    \centering
    \includegraphics[width = 0.4\textwidth]{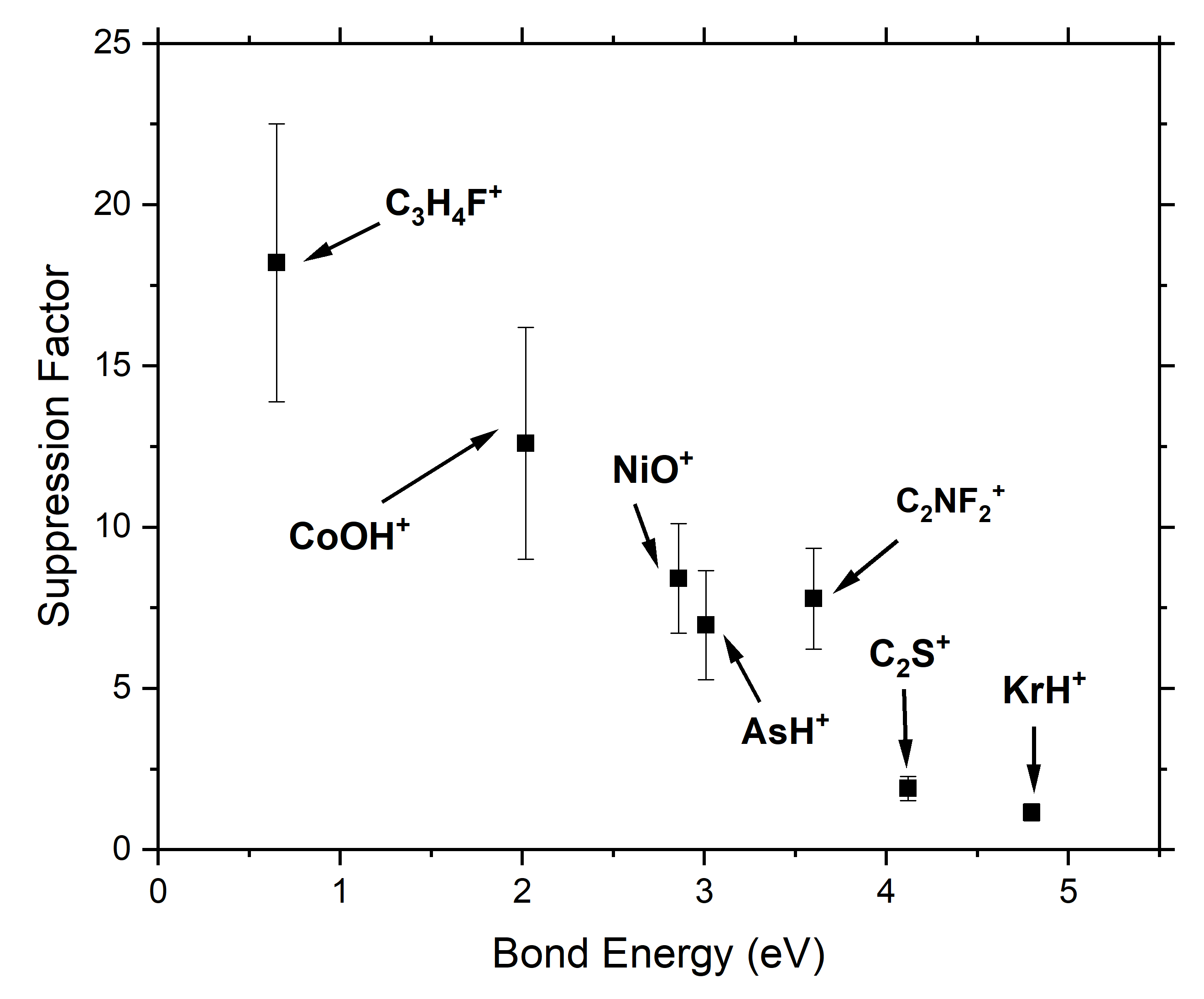}
    \caption{Various molecular species and their bond energies plotted versus the suppression factor that was measured and calculated using Equation \ref{SuppFac}.}
    \label{fig:SF}
\end{figure}

From the data at A/q = 76, a variety of molecular ions with known bond strengths were identified. A plot was generated showing the bond energy \cite{Luo2007ComprehensiveEnergies} versus suppression factor. For polyatomic species, the bond energy of the weakest molecular bond is chosen. Suppression factor is defined as:

\begin{equation}
\label{SuppFac}
    S = \frac{\textrm{Rate}(E_{inj}=10 \textrm{ eV})}{\textrm{Rate}(E_{inj} = 100 \textrm{ eV})} \textrm{,}
\end{equation}

\noindent
and can be considered as the amount by which the transmission is reduced when CID is applied compared to when CID is `turned off'. The results are plotted in Figure \ref{fig:SF}. The relationship provides a clear trend between bond strength and the effect of CID. For more loosely bound molecular ions, a reduction of around an order of magnitude can be expected. However, for strongly bound molecular ions, nearly no reduction in rate for CID occurs. Thus, the gains made from CID can be projected when anticipating various molecular contamination that will be found in the RIB.

\section{Conclusion}

Using a cocktail stable beam of both atomic and molecular ions, CID has been investigated at TITAN. With a better understanding of CID at TITAN, there is now an additional potent tool for cleaning RIB of undesired molecular species. For beams with large amounts of molecular contamination, such as beams from a FEBIAD ion source, CID can be applied by increasing the injection energy from the TITAN RFQ Cooler-Buncher into the Input RFQ of the MR-TOF-MS. This simple change can reduce the molecular contamination by approximately an order of magnitude while accruing no losses of non-molecular species. This additional mode of cleaning will facilitate future mass measurements further away from nuclear stability.

\section*{Acknowledgements}

This work was partially supported by Canadian agencies NSERC and CFI, U.S.A. DOE (grant DE-FG02-93ER40789), Brazils CNPq (grant 249121/2013-1), the Canada-UK Foundation, German institutions DFG (grants FR 601/3-1, SCHE 1969/2-1 and SFB 1245 and through PRISMA Cluster of Excellence), BMBF (grants 05P19RGFN1 and 05P21RGFN1), and by the JLU and GSI under the JLU-GSI strategic Helmholtz partnership agreement.

\label{}

\bibliography{ref}{}

\begin{thebibliography}{10}
\expandafter\ifx\csname url\endcsname\relax
  \def\url#1{\texttt{#1}}\fi
\expandafter\ifx\csname urlprefix\endcsname\relax\def\urlprefix{URL }\fi
\expandafter\ifx\csname href\endcsname\relax
  \def\href#1#2{#2} \def\path#1{#1}\fi

\bibitem{Blumenfeld2013FacilitiesProduction}
Y.~Blumenfeld, T.~Nilsson, P.~Van~Duppen, {Facilities and methods for
  radioactive ion beam production}, Physica Scripta 014023~(T152) (2013).
\newblock \href {https://doi.org/10.1088/0031-8949/2013/T152/014023}
  {\path{doi:10.1088/0031-8949/2013/T152/014023}}.

\bibitem{Hamaker2021}
A.~Hamaker, E.~Leistenschneider, R.~Jain, G.~Bollen, S.~Giuliani, K.~Lund,
  W.~Nazarewicz, L.~Neufcourt, C.~Nicoloff, D.~Puentes, R.~Ringle,
  C.~Sumithrarachchi, I.~Yandow, {Precision mass measurement of lightweight
  self-conjugate nucleus 80Zr}, Nature~(17) (2021) 1408--1412.
\newblock \href {https://doi.org/10.1038/s41567-021-01395-w}
  {\path{doi:10.1038/s41567-021-01395-w}}.

\bibitem{Blaum2004}
K.~Blaum, D.~Beck, G.~Bollen, P.~Delahaye, C.~Gu{\'e}naut, F.~Herfurth,
  A.~Kellerbauer, H.-J. Kluge, D.~Lunney, S.~Schwarz, Population inversion of
  nuclear states by a penning trap mass spectrometer, Europhysics Letters
  67~(4) (2004).

\bibitem{KWIATKOWSKI20159}
A.~Kwiatkowski, G.~Bollen, M.~Redshaw, R.~Ringle, S.~Schwarz, Isobaric beam
  purification for high precision penning trap mass spectrometry of radioactive
  isotope beams with swift, International Journal of Mass Spectrometry 379
  (2015) 9--15.
\newblock \href {https://doi.org/https://doi.org/10.1016/j.ijms.2014.09.016}
  {\path{doi:https://doi.org/10.1016/j.ijms.2014.09.016}}.

\bibitem{Dickel2017IsobarRe-Trapping}
T.~Dickel, W.~R. Pla{\ss}, W.~Lippert, J.~Lang, M.~I. Yavor, H.~Geissel,
  C.~Scheidenberger, {Isobar Separation in a Multiple-Reflection Time-of-Flight
  Mass Spectrometer by Mass-Selective Re-Trapping}, J. Am. Soc. Mass Spectrom.
  28~(6) (2017) 1079--1090.
\newblock \href {https://doi.org/10.1007/s13361-017-1617-z}
  {\path{doi:10.1007/s13361-017-1617-z}}.

\bibitem{MISKUN2021116450}
I.~Miskun, T.~Dickel, S.~A. {San Andrés}, J.~Bergmann, P.~Constantin,
  J.~Ebert, H.~Geissel, F.~Greiner, E.~Haettner, C.~Hornung, W.~Lippert,
  I.~Mardor, I.~Moore, W.~R. Plaß, S.~Purushothaman, A.-K. Rink, M.~P. Reiter,
  C.~Scheidenberger, H.~Weick, Separation of atomic and molecular ions by ion
  mobility with an rf carpet, International Journal of Mass Spectrometry 459
  (2021) 116450.
\newblock \href {https://doi.org/https://doi.org/10.1016/j.ijms.2020.116450}
  {\path{doi:https://doi.org/10.1016/j.ijms.2020.116450}}.

\bibitem{Schury2006BeamCell}
P.~Schury, G.~Bollen, M.~Block, D.~J. Morrissey, R.~Ringle, A.~Prinke,
  J.~Savory, S.~Schwarz, T.~Sun, {Beam purification techniques for low energy
  rare isotope beams from a gas cell}, Hyperfine Interact 173 (2006) 165--170.
\newblock \href {https://doi.org/10.1007/s10751-007-9553-0}
  {\path{doi:10.1007/s10751-007-9553-0}}.

\bibitem{Greiner2020RemovalMethod}
F.~Greiner, T.~Dickel, S.~Ayet San~Andr{\'{e}}s, J.~Bergmann, P.~Constantin,
  J.~Ebert, H.~Geissel, E.~Haettner, C.~Hornung, I.~Miskun, W.~Lippert,
  I.~Mardor, I.~Moore, W.~R. Pla{\ss}, S.~Purushothaman, A.~K. Rink, M.~P.
  Reiter, C.~Scheidenberger, H.~Weick, {Removal of molecular contamination in
  low-energy RIBs by the isolation-dissociation-isolation method}, Nuclear
  Instruments and Methods in Physics Research, Section B: Beam Interactions
  with Materials and Atoms 463 (2020) 324--326.
\newblock \href {https://doi.org/10.1016/j.nimb.2019.04.072}
  {\path{doi:10.1016/j.nimb.2019.04.072}}.

\bibitem{Dilling2003TheIsotopes}
J.~Dilling, P.~Bricault, M.~Smith, H.~J. Kluge, {The proposed TITAN facility at
  ISAC for very precise mass measurements on highly charged short-lived
  isotopes}, Nucl. Instrum. Meth. Phys. Res. B 204 (2003) 492--496.
\newblock \href {https://doi.org/10.1016/S0168-583X(02)02118-3}
  {\path{doi:10.1016/S0168-583X(02)02118-3}}.

\bibitem{Ball2016TheBeams}
G.~C. Ball, G.~Hackman, R.~Kr{\"{u}}cken, {The TRIUMF-ISAC facility: two
  decades of discovery with rare isotope beams}, Physica Scripta 91~(093002)
  (2016).
\newblock \href {https://doi.org/10.1088/0031-8949/91/9/093002}
  {\path{doi:10.1088/0031-8949/91/9/093002}}.

\bibitem{Brunner2012TITANsPerformance}
T.~Brunner, M.~J. Smith, M.~Brodeur, S.~Ettenauer, A.~T. Gallant, V.~V. Simon,
  A.~Chaudhuri, A.~Lapierre, E.~Man{\'{e}}, R.~Ringle, M.~C. Simon, J.~A. Vaz,
  P.~Delheij, M.~Good, M.~R. Pearson, J.~Dilling, {TITAN's digital RFQ ion beam
  cooler and buncher, operation and performance}, Nucl. Instrum. Meth. Phys.
  Res. A 676 (2012) 32--43.
\newblock \href {https://doi.org/10.1016/j.nima.2012.02.004}
  {\path{doi:10.1016/j.nima.2012.02.004}}.

\bibitem{Jesch2015TheTRIUMF}
C.~Jesch, T.~Dickel, W.~R. Pla{\ss}, D.~Short, S.~A. San~Andres, J.~Dilling,
  H.~Geissel, F.~Greiner, J.~Lang, K.~G. Leach, W.~Lippert, C.~Scheidenberger,
  M.~I. Yavor, {The MR-TOF-MS isobar separator for the TITAN facility at
  TRIUMF}, Hyperfine Interact 235 (2015) 97--106.
\newblock \href {https://doi.org/10.1007/s10751-015-1184-2}
  {\path{doi:10.1007/s10751-015-1184-2}}.

\bibitem{Leistenschneider2018DawningIsotopes}
E.~Leistenschneider, M.~P. Reiter, S.~A. San~Andr{\'{e}}s, B.~Kootte, J.~D.
  Holt, P.~Navr{\'{a}}til, C.~Babcock, C.~Barbieri, B.~R. Barquest,
  J.~Bergmann, J.~Bollig, T.~Brunner, E.~Dunling, A.~Finlay, H.~Geissel,
  L.~Graham, F.~Greiner, H.~Hergert, C.~Hornung, C.~Jesch, R.~Klawitter,
  Y.~Lan, D.~Lascar, K.~G. Leach, W.~Lippert, J.~E. McKay, S.~F. Paul,
  A.~Schwenk, D.~Short, J.~Simonis, V.~Som{\`{a}}, R.~Steinbr{\"{u}}gge, S.~R.
  Stroberg, R.~Thompson, M.~E. Wieser, C.~Will, M.~I. Yavor, C.~Andreoiu,
  T.~Dickel, I.~Dillmann, G.~Gwinner, W.~R. Pla{\ss}, C.~Scheidenberger, A.~A.
  Kwiatkowski, J.~Dilling, {Dawning of the N = 32 Shell Closure Seen through
  Precision Mass Measurements of Neutron-Rich Titanium Isotopes}, Phys. Rev.
  Lett. 120~(6) (2018) 62503.
\newblock \href {https://doi.org/10.1103/PhysRevLett.120.062503}
  {\path{doi:10.1103/PhysRevLett.120.062503}}.

\bibitem{Reiter2018QuenchingIsotopes}
M.~P. Reiter, S.~A. San~Andr{\'{e}}s, E.~Dunling, B.~Kootte,
  E.~Leistenschneider, C.~Andreoiu, C.~Babcock, B.~R. Barquest, J.~Bollig,
  T.~Brunner, I.~Dillmann, A.~Finlay, G.~Gwinner, L.~Graham, J.~D. Holt,
  C.~Hornung, C.~Jesch, R.~Klawitter, Y.~Lan, D.~Lascar, J.~E. McKay, S.~F.
  Paul, R.~Steinbr{\"{u}}gge, R.~Thompson, J.~L. Tracy, M.~E. Wieser, C.~Will,
  T.~Dickel, W.~R. Pla{\ss}, C.~Scheidenberger, A.~A. Kwiatkowski, J.~Dilling,
  {Quenching of the N=32 neutron shell closure studied via precision mass
  measurements of neutron-rich vanadium isotopes}, Phys. Rev. C 98~(2) (2018)
  024310.
\newblock \href {https://doi.org/10.1103/PhysRevC.98.024310}
  {\path{doi:10.1103/PhysRevC.98.024310}}.

\bibitem{Beck2021MassShell}
S.~Beck, B.~Kootte, I.~Dedes, T.~Dickel, A.~A. Kwiatkowski, E.~M.
  Lykiardopoulou, W.~R. Pla{\ss}, M.~P. Reiter, C.~Andreoiu, J.~Bergmann,
  T.~Brunner, D.~Curien, J.~Dilling, J.~Dudek, E.~Dunling, J.~Flowerdew,
  A.~Gaamouci, L.~Graham, G.~Gwinner, A.~Jacobs, R.~Klawitter, Y.~Lan,
  E.~Leistenschneider, N.~Minkov, V.~Monier, I.~Mukul, S.~F. Paul,
  C.~Scheidenberger, R.~I. Thompson, J.~L. Tracy, M.~Vansteenkiste, H.~L. Wang,
  M.~E. Wieser, C.~Will, J.~Yang, {Mass Measurements of Neutron-Deficient Yb
  Isotopes and Nuclear Structure at the Extreme Proton-Rich Side of the N=82
  Shell}, Physical Review Letters 127~(11) (2021) 112501.
\newblock \href {https://doi.org/10.1103/PhysRevLett.127.112501}
  {\path{doi:10.1103/PhysRevLett.127.112501}}.

\bibitem{Mukul2021}
I.~Mukul, C.~Andreoiu, J.~Bergmann, M.~Brodeur, T.~Brunner, K.~A. Dietrich,
  T.~Dickel, I.~Dillmann, E.~Dunling, D.~Fusco, G.~Gwinner, C.~Izzo, A.~Jacobs,
  B.~Kootte, Y.~Lan, E.~Leistenschneider, E.~M. Lykiardopoulou, S.~F. Paul,
  M.~P. Reiter, J.~L. Tracy, J.~Dilling, A.~A. Kwiatkowski, Examining the
  nuclear mass surface of rb and sr isotopes in the $a\ensuremath{\approx}104$
  region via precision mass measurements, Phys. Rev. C 103 (2021) 044320.
\newblock \href {https://doi.org/10.1103/PhysRevC.103.044320}
  {\path{doi:10.1103/PhysRevC.103.044320}}.

\bibitem{Izzo2021}
C.~Izzo, J.~Bergmann, K.~A. Dietrich, E.~Dunling, D.~Fusco, A.~Jacobs,
  B.~Kootte, G.~Kripk\'o-Koncz, Y.~Lan, E.~Leistenschneider, E.~M.
  Lykiardopoulou, I.~Mukul, S.~F. Paul, M.~P. Reiter, J.~L. Tracy, C.~Andreoiu,
  T.~Brunner, T.~Dickel, J.~Dilling, I.~Dillmann, G.~Gwinner, D.~Lascar, K.~G.
  Leach, W.~R. Pla\ss{}, C.~Scheidenberger, M.~E. Wieser, A.~A. Kwiatkowski,
  Mass measurements of neutron-rich indium isotopes for $r$-process studies,
  Phys. Rev. C 103 (2021) 025811.
\newblock \href {https://doi.org/10.1103/PhysRevC.103.025811}
  {\path{doi:10.1103/PhysRevC.103.025811}}.

\bibitem{Reiter2020ImprovedMR-TOF-MS}
M.~P. Reiter, F.~Ames, C.~Andreoiu, S.~Ayet San~Andr{\'{e}}s, C.~Babcock, B.~R.
  Barquest, J.~Bergmann, J.~Bollig, T.~Brunner, T.~Dickel, J.~Dilling,
  I.~Dillmann, E.~Dunling, A.~Finlay, G.~Gwinner, L.~Graham, C.~Hornung,
  B.~Kootte, R.~Klawitter, P.~Kunz, Y.~Lan, D.~Lascar, J.~Lassen,
  E.~Leistenschneider, R.~Li, J.~E. McKay, M.~Mostamand, S.~F. Paul, W.~R.
  Pla{\ss}, C.~Scheidenberger, B.~E. Schultz, R.~Steinbr{\"{u}}gge,
  A.~Teigelhoefer, R.~Thompson, M.~E. Wieser, C.~Will, A.~A. Kwiatkowski,
  {Improved beam diagnostics and optimization at ISAC via TITAN's MR-TOF-MS},
  Nuclear Instruments and Methods in Physics Research, Section B: Beam
  Interactions with Materials and Atoms 463 (2020) 431--436.
\newblock \href {https://doi.org/10.1016/j.nimb.2019.04.034}
  {\path{doi:10.1016/j.nimb.2019.04.034}}.

\bibitem{Pla2008IsobarFacilities}
W.~R. Pla{\ss}, T.~Dickel, U.~Czok, H.~Geissel, M.~Petrick, K.~Reinheimer,
  C.~Scheidenberger, M.~I. Yavor, {Isobar separation by time-of-flight mass
  spectrometry for low-energy radioactive ion beam facilities}, Nucl. Instrum.
  Meth. Phys. Res. B 266 (2008) 4560--4564.
\newblock \href {https://doi.org/10.1016/j.nimb.2008.05.079}
  {\path{doi:10.1016/j.nimb.2008.05.079}}.

\bibitem{Dickel2015ANuclei}
T.~Dickel, W.~R. Pla{\ss}, A.~Becker, U.~Czok, H.~Geissel, E.~Haettner,
  C.~Jesch, W.~Kinsel, M.~Petrick, C.~Scheidenberger, A.~Simon, M.~I. Yavor, {A
  high-performance multiple-reflection time-of-flight mass spectrometer and
  isobar separator for the research with exotic nuclei}, Nucl. Instrum. Meth.
  Phys. Res. A 777 (2015) 172--188.
\newblock \href {https://doi.org/10.1016/j.nima.2014.12.094}
  {\path{doi:10.1016/j.nima.2014.12.094}}.

\bibitem{Pla2015High-performanceSpectrometry}
W.~R. Pla{\ss}, T.~Dickel, S.~San~Andres, J.~Ebert, F.~Greiner, C.~Hornung,
  C.~Jesch, J.~Lang, W.~Lippert, T.~Majoros, D.~Short, H.~Geissel, E.~Haettner,
  M.~P. Reiter, A.~K. Rink, C.~Scheidenberger, M.~I. Yavor, {High-performance
  multiple-reflection time-of-flight mass spectrometers for research with
  exotic nuclei and for analytical mass spectrometry}, Physica Scripta
  2015~(T166) (2015).
\newblock \href {https://doi.org/10.1088/0031-8949/2015/T166/014069}
  {\path{doi:10.1088/0031-8949/2015/T166/014069}}.

\bibitem{Yavor2015Ion-opticalSeparator}
M.~I. Yavor, W.~R. Pla{\ss}, T.~Dickel, H.~Geissel, C.~Scheidenberger,
  {Ion-optical design of a high-performance multiple-reflection time-of-flight
  mass spectrometer and isobar separator}, International Journal of Mass
  Spectrometry (2015).
\newblock \href {https://doi.org/10.1016/j.ijms.2015.01.002}
  {\path{doi:10.1016/j.ijms.2015.01.002}}.

\bibitem{AndresThesis}
S.~Andres, Developments for multiple-reflection time-of-flight mass
  spectrometers and their application to high-resolution accurate mass
  measurements of short-lived exotic nuclei, Ph.D. thesis,
  Justus-Liebig-Universit\"{a}t Gie{\ss}en (2018).

\bibitem{Dickel2017DynamicalSpectrometers}
T.~Dickel, M.~I. Yavor, J.~Lang, W.~R. Pla{\ss}, W.~Lippert, H.~Geissel,
  C.~Scheidenberger, {Dynamical time focus shift in multiple-reflection
  time-of-flight mass spectrometers}, International Journal of Mass
  Spectrometry 412 (2017).
\newblock \href {https://doi.org/10.1016/j.ijms.2016.11.005}
  {\path{doi:10.1016/j.ijms.2016.11.005}}.

\bibitem{Paul2021}
S.~F. Paul, J.~Bergmann, J.~D. Cardona, K.~A. Dietrich, E.~Dunling,
  Z.~Hockenbery, C.~Hornung, C.~Izzo, A.~Jacobs, A.~Javaji, B.~Kootte, Y.~Lan,
  E.~Leistenschneider, E.~M. Lykiardopoulou, I.~Mukul, T.~Murb\"ock, W.~S.
  Porter, R.~Silwal, M.~B. Smith, J.~Ringuette, T.~Brunner, T.~Dickel,
  I.~Dillmann, G.~Gwinner, M.~MacCormick, M.~P. Reiter, H.~Schatz, N.~A.
  Smirnova, J.~Dilling, A.~A. Kwiatkowski, Mass measurements of
  $^{60--63}\mathrm{Ga}$ reduce x-ray burst model uncertainties and extend the
  evaluated $t=1$ isobaric multiplet mass equation, Phys. Rev. C 104 (2021)
  065803.
\newblock \href {https://doi.org/10.1103/PhysRevC.104.065803}
  {\path{doi:10.1103/PhysRevC.104.065803}}.

\bibitem{Porter2022}
W.~S. Porter, B.~Ashrafkhani, J.~Bergmann, C.~Brown, T.~Brunner, J.~D. Cardona,
  D.~Curien, I.~Dedes, T.~Dickel, J.~Dudek, E.~Dunling, G.~Gwinner,
  Z.~Hockenbery, J.~D. Holt, C.~Hornung, C.~Izzo, A.~Jacobs, A.~Javaji,
  B.~Kootte, G.~Kripk\'o-Koncz, E.~M. Lykiardopoulou, T.~Miyagi, I.~Mukul,
  T.~Murb\"ock, W.~R. Pla\ss{}, M.~P. Reiter, J.~Ringuette, C.~Scheidenberger,
  R.~Silwal, C.~Walls, H.~L. Wang, Y.~Wang, J.~Yang, J.~Dilling, A.~A.
  Kwiatkowski, Mapping the $n=40$ island of inversion: Precision mass
  measurements of neutron-rich fe isotopes, Phys. Rev. C 105 (2022) L041301.
\newblock \href {https://doi.org/10.1103/PhysRevC.105.L041301}
  {\path{doi:10.1103/PhysRevC.105.L041301}}.

\bibitem{REITER2021165823}
M.~Reiter, S.~A.~S. Andrés, J.~Bergmann, T.~Dickel, J.~Dilling, A.~Jacobs,
  A.~Kwiatkowski, W.~Plaß, C.~Scheidenberger, D.~Short, C.~Will, C.~Babcock,
  E.~Dunling, A.~Finlay, C.~Hornung, C.~Jesch, R.~Klawitter, B.~Kootte,
  D.~Lascar, E.~Leistenschneider, T.~Murböck, S.~Paul, M.~Yavor, Commissioning
  and performance of titan’s multiple-reflection time-of-flight
  mass-spectrometer and isobar separator, Nuclear Instruments and Methods in
  Physics Research Section A: Accelerators, Spectrometers, Detectors and
  Associated Equipment 1018 (2021) 165823.
\newblock \href {https://doi.org/https://doi.org/10.1016/j.nima.2021.165823}
  {\path{doi:https://doi.org/10.1016/j.nima.2021.165823}}.

\bibitem{McLuckey1992PrinciplesSpectrometry}
S.~A. McLuckey, {Principles of collisional activation in analytical mass
  spectrometry}, J. Am. Soc. Mass Spectrom. 3 (1992) 599--614.
\newblock \href {https://doi.org/10.1016/1044-0305(92)85001-Z}
  {\path{doi:10.1016/1044-0305(92)85001-Z}}.

\bibitem{Beynon1973DesignSpectrometer}
J.~H. Beynon, R.~G. Cooks, J.~W. Amy, W.~E. Baitinger, T.~Y. Ridley, {Design
  and Performance of a Mass-analyzed Ion Kinetic Energy (MIKE) Spectrometer},
  Anal. Chem. 45~(12) (1973) 1023A--1031A.
\newblock \href {https://doi.org/10.1021/ac60334a763}
  {\path{doi:10.1021/ac60334a763}}.

\bibitem{Morris1996HighSpectrometer}
H.~R. Morris, T.~Paxton, A.~Dell, J.~Langhorne, M.~Berg, R.~S. Bordoli,
  J.~Hoyes, R.~H. Bateman, {High Sensitivity Collisionally-activated
  Decomposition Tandem Mass Spectrometry on a Novel
  Quadrupole/Orthogonal-acceleration Time-of-flight Mass Spectrometer}, Rapid
  Communications in Mass Spectrometry 10~(8) (1996) 889--896.
\newblock \href
  {https://doi.org/10.1002/(SICI)1097-0231(19960610)10:8$<$889::AID-RCM615$>$3.0.CO;2-F}
  {\path{doi:10.1002/(SICI)1097-0231(19960610)10:8$<$889::AID-RCM615$>$3.0.CO;2-F}}.

\bibitem{Louris1987InstrumentationSpectrometry}
J.~N. Louris, R.~G. Cooks, J.~E. Syka, P.~E. Kelley, G.~C. Stafford, J.~F.
  Todd, {Instrumentation, Applications, and Energy Deposition in Quadrupole
  Ion-Trap Tandem Mass Spectrometry}, Anal. Chem. 59~(13) (1987) 1677--1685.
\newblock \href {https://doi.org/10.1021/ac00140a021}
  {\path{doi:10.1021/ac00140a021}}.

\bibitem{Wells2005Collision-inducedProteins}
J.~M. Wells, S.~A. McLuckey, {Collision-induced dissociation (CID) of peptides
  and proteins}, Methods Enzym. 402 (2005) 148--185.
\newblock \href {https://doi.org/10.1016/S0076-6879(05)02005-7}
  {\path{doi:10.1016/S0076-6879(05)02005-7}}.

\bibitem{Boyd1984Angle-DependenceCollisions}
R.~K. Boyd, E.~E. Kingston, A.~G. Brenton, J.~H. Beynon, {Angle-Dependence of
  Ion Kinetic Energy Spectra Obtained by Using Mass Spectrometers. I.
  Theoretical Consequences of Conservation Laws for Collisions}, Proc. R. Soc.
  Lond. A 392~(1802) (1984) 59--88.
\newblock \href {https://doi.org/10.1098/rspa.1984.0024}
  {\path{doi:10.1098/rspa.1984.0024}}.

\bibitem{Douglas1982MechanismSpectrometry}
D.~J. Douglas, {Mechanism of the collision-induced dissociation of polyatomic
  ions studied by triple quadrupole mass spectrometry}, J. Phys. Chem 86~(2)
  (1982) 185--191.
\newblock \href {https://doi.org/10.1021/j100391a011}
  {\path{doi:10.1021/j100391a011}}.

\bibitem{Luo2007ComprehensiveEnergies}
Y.-R. Luo, {Comprehensive Handbook of Chemical Bond Energies}, CRC Press,
  Taylor and Francis Group, Boca Raton, 2007.

\end{thebibliography}

\bibliographystyle{elsarticle-num}

\end{document}